\documentclass{ws-brl}
\usepackage{graphicx,xspace,units,subfigure}
\usepackage{float}
\usepackage{amsmath}
\usepackage{amssymb}
\usepackage{mathrsfs}
\usepackage[normalem]{ulem}
\usepackage{epsfig}

\newcommand{\noi}{\noindent}
\newcommand{\be}{\begin{equation}}
\newcommand{\ee}{\end{equation}}

\begin{document}

\markboth{Alessandro Taloni, Fabio Marchesoni}
{Interacting single-file system: Fractional Langevin formulation versus diffusion-noise approach}

\catchline{}{}{}{}{}

\title{Interacting single-file system: Fractional Langevin formulation versus diffusion-noise approach}

\author{ALESSANDRO TALONI}
\address{
CNR-IENI, Via R. Cozzi 53\\  
Milano, 20125, Italy\\
alessandro.taloni@gmail.com}

\author{FABIO MARCHESONI}
\address{Dipartimento di Fisica, Universit\'a di Camerino,Via Madonna delle Carceri, 9\\
Camerino, I-62032, Italy\\
fabio.marchesoni@pg.infn.it}

\maketitle
\begin{history}
\received{Day Month Year}
\revised{Day Month Year}
\end{history}

\begin{abstract}

We review the latest advances in the analytical modelling of single
file diffusion. We focus first on the derivation of the
fractional Langevin equation that describes the motion of a tagged
file particle. We then propose an alternative derivation of the very
same stochastic equation by starting from the diffusion-noise
formalism for the time evolution of the file density.
\end{abstract}

\maketitle
\section{Introduction}

The single file model describes particle diffusion through channels
or pores. Firstly introduced in the biological context \cite{Hodgkin_JPhysiol_1955}
\cite{Harris_JApplProb_1965}, single file diffusion (SFD) gained
widespread interest in the mathematical physical community
\cite{Fedders_PRB_1978}\cite{Kollmann_PRL_2003}\cite{Alexander_PRB_1978},
\cite{Karger_PRA_1992}, \cite{Arratia_AnnProb_1983}, \cite{Percus_PRA_1974},
\cite{Levitt_PRA_1973}, as a simple, analytically tractable model
demonstrating anomalous diffusion. Due to the simplicity of its
formulation, the limited number of the parameters characterizing its
dynamics, and its immediate relation to real physical phenomena, such
a stylized model provides an ideal benchmark against which we can
test our understanding of anomalous diffusion.

The single file model is defined as an assembly of identical
interacting particles diffusing on a one dimensional (1D) substrate.
The particles are non-passing, i.e. they retain their initial file
ordering at any time, a spatial constraint which goes under the name
of \emph{single filing condition}. This means that one can pick any
file particle and attempt to describe its stochastic dynamics; this
particle is usually referred to as the \emph{tagged particle}. It has
been known since the very first formulation of the model
\cite{Harris_JApplProb_1965}, that the mean square displacement of
the tagged particle undergoes the asymptotic law

\be \langle \left[x_n(t)-x_n(0)\right]^2\rangle =
2\sqrt{\frac{Dt}{\pi\langle\rho\rangle^2}}, \label{SF_diffusion}
\ee

\noi where $D$ is the diffusion coefficient of the freely diffusing
particles and $\langle\rho\rangle$ denotes the average equilibrium
file density.

In this paper we review some of the major advances in the analytical
treatment of single file systems achieved over the last decade.
First, we focus on the action of an external potential and how it
affects the tagged particle dynamics. Second, we discuss two
alternative descriptions of the SFD: one, called fractional Langevin
approach, allows to extract an effective stochastic equation for the
tagged particle from the detailed file dynamics. The other one,
termed diffusion-noise approach, introduces an intermediate step,
namely, the equation for time evolution of the file density; the
fractional Langevin equation of the tagged particle is then derived
by solving the relevant density equation.

\subsection{Interacting single file system}

Let us consider a 1D file of $N$ pointlike Brownian particles moving
on a ring according to the Langevin equations

\be \frac{d^2x_n(t)}{dt^2}=-\gamma
\frac{dx_n(t)}{dt}-\frac{\partial}{\partial
  x_n}U\left(x_1,\dots,x_N;t\right)+\eta_n(t),
\label{LE}
\ee

\noi with $n=1,\dots,N$,  where the damping constant $\gamma$ and the
noises $\eta_n(t)$ satisfy the fluctuation-dissipation relations

\be
\langle\eta_i(t)\eta_j(t')\rangle=2k_BT\gamma\delta_{i,j}\delta(t-t')
\label{FD}. \ee

\noi where $\delta$ stands for the Dirac's delta function. The file
particles interact with both their nearest neighbors and the
environment. The potential function in Eq. (\ref{LE}) consists of two
contributions,

\be
U\left(x_1,\dots,x_N;t\right)=\sum_{n,m=1}
U_{HC}\left(\left|x_n-x_m\right|\right)+U_{\rm
int}\left(x_1,\dots,x_N;t\right), 
\label{potential} 
\ee

\noi with the hard-core repulsion

\be
U_{HC}\left(\left|x_n-x_m\right|\right)=
\left\{
\begin{array}{ccc}
\infty  & & |x_n-x_m|=0\\
0       & & {\rm otherwise},
\end{array}
\right.
\label{potential_HC}
\ee

\noi establishing the filing condition. As stated at the section beginning in the following we will focus on the zero  particles' size limit, however it must be stressed that the truly interaction potential for particles with a finite size $a$ yields $U_{HC}\left(\left|x_n-x_m\right|\right)=\infty$ if $|x_n-x_m|=a$.  The additional
substrate potential $U_{\rm int}$, is often introduced to model
distinct physical situations.

\begin{figure}[ph]
\centerline{\psfig{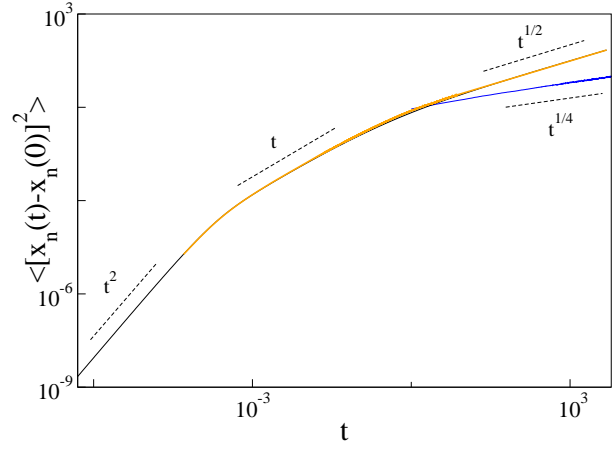}}
\vspace*{8pt}
\caption{Tagged-particle mean-square displacement
curve (solid orange) showing the three characteristic diffusive
regimes: ballistic, diffusive, and subdiffusive. The corresponding
theoretical prediction (solid black curve) was obtained by
numerically integrating Eq.(\ref{GLE}). Integration parameters are
$\langle\rho\rangle=0.1$, $k_BT=1.0$, and $\gamma=0.1$. The
subdiffusive behavior of a tagged particle in a single file of
subdiffusing particles, $\langle\left[x_{\rm free}(t)-x_{\rm
free}(0)\right]^2\rangle\propto t^{\alpha}$ with $\alpha=0.5$, see
Eq. (\ref{percus-rule}), is shown as a test of the one-half rule.
\label{fig1}}
\end{figure}

\begin{itemize}

\item \emph{File interaction with a substrate.} A typical example is represented by
non-passing particles confined to constrained geometries, like
compartmentalized narrow channels \cite{burada_ChemPhysChem_2009}.  In
this case the $U_{\rm int}$ maxima model the entropic barriers
opposing the particle diffusion through the compartment pores. The
most general form of such interaction potential is given by

\be
U_{\rm int}\left(x_1,\dots,x_N;t\right)= \int dX\,\sum_{n=1}
U(X;t)\,\delta\left(X-x_n\right).
\label{int_substrate}
\ee

\noi Note that the spatial
coordinate, $X$, has been capitalized so as to be distinguished from
the particle trajectories, $x_1(t), \dots, x_N(t)$. A substantial simplification comes from the assumption that the
potential is \emph{time-independent}, i.e $U(X;t)\equiv U(X)$. The
first example of substrate po[tential energy reported in the
literature,

\be U(X)=d\left[1-\cos\left(\frac{2\pi X}{l}\right)\right],
\label{U_sinusoidal} \ee

\noi modeled the periodic corrugations of a narrow quasi 1D channel
~\cite{Taloni_PRL_2006}. In Eq.(\ref{U_sinusoidal}) $d$ and $l$
denote respectively the amplitude and period of the channel wall
corrugation. The authors of Refs.~\cite{Barkai_PRL_2009}\cite{Barkai_PRE_2010} established a general expression for the tagged
particle diffusion in mirror symmetric potentials, which they were
able to compute explicitly only for the harmonic potential

\be U(X)= \frac{\omega^2}{2} X^2. \label{U_harmonic} \ee

\noi Their results corroborated, both numerically and theoretically,
the validity of the so-called Percus diffusion formula
~\cite{Percus_PRA_1974},

\be \langle \left[x_n(t)-x_n(0)\right]^2\rangle =
\frac{\langle\left|x_{\rm free}(t)-x_{\rm
free}(0)\right|\rangle}{\langle\rho\rangle}, \label{percus-rule} \ee

\noi where $x_{\rm free}(t)$ represents the trajectory of a single
(passing) particle under the conditions given. The Percus' rule holds whenever the mean square displacement is described in the restframe of the center of mass. In fact, Percus'
formula applies to any kind of single particle dynamics, namely, to
Brownian particles, like in Eq.(\ref{LE}), as well as to anomalous
diffusing particles \cite{Barkai_PRE_2010}\cite{Bandyopadhyay_EPL_2008}\cite{Flomenbom_EPL_2008}\cite{Marchesoni_PRB_1998}\cite{Flomenbom_epl_2011}\cite{Flomenbom_PRE_2010}\cite{Flomenbom_Phys_Lett_2010}. For instance, if the free
diffusion law of a single particle is

\be \langle \left[x_{\rm free}(t)-x_{\rm free}(0)\right]^2\rangle
\propto t^\alpha \label{free_anomalous}, \ee

\noi with $\alpha>0$, then, from Eq.(\ref{percus-rule}) one obtains
$\langle \left[x_n(t)-x_n(0)\right]^2\rangle\propto
t^{\alpha/2}/\langle\rho\rangle $. For this reason Percus' relation
(\ref{percus-rule}) is also known as the ``one-half rule'' (see
Fig.\ref{fig1}). Finally, we mention that, as numerically proven in
Ref.\cite{Ben-Naim_PRL_2009}, the presence of weak substrate disorder
may  induce superdiffusive SFD corresponding to $\alpha
>1$.

\item \emph{One-particle interaction}. This type of interaction potential,
\be U_{\rm int}\left(x_1,\dots,x_N;t\right) =\int dX\,
U(X;t)\,\delta\left(X-x_n\right), \label{int_singlepart} \ee

\noi was introduced to mimic the effects of an external field acting
on one particle, alone ~\cite{Taloni_PRE_2011}. In this case, we
agree to identify the tagged particle as the particle directly
coupled to the potential. Popular examples are provided by optical or
magnetic tweezers, which trap a single particle in a harmonic
potential $U(X;t)\equiv \omega^2{X^2}/{2}$ with characteristic
frequency $\omega$, while leaving the rest of the file unaffected.
Moreover, if the tagged particle carries an electric charge, $q$,
applying a constant external electric field, $E$, results in the
one-particle potential $U(X;t)=-FX$ with $F=qE$. In such a case and,
more generally, for any constant field of force $F$, the ensuing
drift of the tagged particle $\langle \left[x_n(t)-x_n(0)\right]\rangle_F$ fulfills the celebrated Einstein
relation in the form ~\cite{Burlatsky_PRE_1996}, \cite{Taloni_PRE_2008},
\cite{Taloni_PRE_2011},
\cite{Taloni_MNP_2013}\cite{Leibovich_PRE_2014}\cite{Villamaina_JSTAT_2008}\cite{Illien_PRL_2013}:

\be \langle \left[x_n(t)-x_n(0)\right]\rangle_F = F\frac{\langle
\left[x_n(t)-x_n(0)\right]^2\rangle }{2k_BT}. \label{Einstein_rel}
\ee

\noi Another kind of external interaction potential addressed in the
recent literature, is the sinusoidal time-dependent drive $U(X;t) =-
AX\cos(\omega t)$ ~\cite{Taloni_PRE_2008}, \cite{Taloni_PRE_2011}. Under
these conditions the Green-Kubo relation, for small enough $A$, still holds rigorously for
the tagged particle,

\be \langle v_n(\omega)v_n(-\omega)\rangle = 4\pi
k_BT\frac{A}{\omega} Re\left[\mu_n(\omega)\right], \label{green-kubo}
\ee

\noi where $v\equiv \dot{x}$ and $\mu_n(\omega)$ is the complex
mobility of the tagged particle  defined via the identity $\langle
v_n(t)\rangle = Re\left[\mu_n(\omega)Ae^{i\omega
t}\right]$. However, the time modulation applied to the tagged
particle affects, through the hard-core collisions, the dynamics of
all remaining file constituents. In Refs.\cite{Taloni_PRE_2011}\cite{Taloni_MNP_2013} the drift and velocity correlation functions of all
untagged file particle were proven to fulfill generalized Kubo and
Green-Kubo relations obtained as extensions, respectively, of
Eqs.(\ref{Einstein_rel}) and (\ref{green-kubo}). Recently the same approach has been extended to a file system with distributed friction constants \cite{Lomholt_PRE_2014}.

\item \emph{File particle-particles interaction}. In this case no
external field affects the file dynamics, while the particles are
supposed to interact through some additional finite-range binary
potential

\be
 U_{\rm int}\left(x_1,\dots,x_N;X,t\right) =  \sum_{n,m=1} U_{\rm
int}\left(\left|x_n-x_m\right|\right)
 \label{int_file_part} 
\ee

\noi This kind of potential is particularly suited to describe real
physical situations, where the single file particles are taken to be
electrically or magnetically charged. As a matter of fact, several
experiments have been carried out using particles with magnetic
~\cite{Wei_Science_2000}, electric dipole \cite{Lin_EPL_2002}\cite{Lutz_J_Phys_Cond_Matt_2004}\cite{Lutz_PRL_2004}, or screened electrostatic
pair interactions \cite{Coupier_PRE_2006}\cite{Coupier_EPL_2007},
\cite{Coste_PRE_2010}, \cite{Delfau_PRE_2010}.
These  experimental studies provided strong evidence of the
subdiffusive asymptotic behavior predicted in
Eq.(\ref{SF_diffusion}), despite an apparent slowing down in the case
of screened electrostatic interactions \cite{Coupier_EPL_2007}. An exact analytical result
was obtained for the SFD of tagged particles interacting through the
quadratic potential

\be U_{\rm int}\left(\left|x_n-x_m\right|\right)=
\langle\rho\rangle^z\frac{k_z}{4}A_{n,m}^z\left(x_n-x_m\right)^2,
\label{potential_LR} \ee

\noi where $k_z$ is a constant,

\be A_{n,m}^z=\frac{\Gamma\left(\left|n-m\right|-\frac{z}{2}\right)
\Gamma\left(z+1\right)}{\pi\Gamma\left(\left|n-m\right|+\frac{z}{2}
+1\right)}\sin\left(\frac{z\pi}{2}\right), \label{toepliz} \ee

\noi is a Toeplitz matrix \cite{Zoia_PRE_2007}, and $\Gamma(x)$
denotes a Gamma function with argument $x$. An explicit calculation
shows that for $1<z<2$ the SFD formula of Eq.(\ref{SF_diffusion})
must be replaced by $\langle \left[x_{n}(t)-x_{n}(0)\right]^2\rangle
\propto t^{\frac{z-1}{z}} $ \cite{Taloni-unpublished}.

\end{itemize}

\section{Langevin equation for the tagged particle}

The question whether the tagged particle trajectory can be described
by a {\it bona fide} Langevin equation upon integrating out all
remaining file variables, was first addressed in Ref.
\cite{Taloni_PRE_2008}. According to these authors the hard-core
interactions of the tagged particle with its nearest neighbors was
well reproduced by two phenomenological terms, to be regarded
respectively as an intrinsic noise and a file damping term connected
by a fluctuation-dissipation relation of the sort. The resulting
phenomenological Langevin equation for the single file of
Eqs.(\ref{LE}) and (\ref{potential}) with $U_{\rm int}=0$, reads

\be
\frac{d^2x_n(t)}{dt^2} + \gamma \frac{dx_n(t)}{dt}+
2\sqrt{\gamma\,k_BT}\langle\rho\rangle\frac{d^{1/2}x_n(t)}{dt^{1/2}} =\xi_n(t),
\label{GLE}
\ee

\noi where the Caputo fractional derivative is defined by
\cite{Samko,Podlubny}

\be
\frac{d^{1/2}f(t)}{dt^{1/2}}=\frac{1}{\Gamma\left(\frac{1}{2}\right)}\int_0^t\frac{df(t')/dt'}{\left|t-t'\right|^{\frac{1}{2}}}\,dt'.
\label{caputo_der}
\ee

\noi Such a fractional Langevin equation (FLE) has been validated by
extensive numerical simulations
\cite{Taloni_PRE_2008}\cite{Lizana_PRE_2010}. Moreover, it can be easily
rewritten in the form of a generalized Langevin equation (GLE)
~\cite{Mori_1956}\cite{Mori_1965_first}\cite{Mori_1965_second}\cite{Kubo_1966}

\be
\begin{array}{l}
\dot{x}_n(t)  =  v_n(t)\\
\dot{v}_n(t)  = -\int_0^t\kappa\left(t-t'\right)v(t')\,dt'+\xi_n(t)
\label{GLE_velocity}
\end{array}
\ee

\noi by introducing the phenomenological damping kernel
$\kappa(t)=2\,\gamma\left[\delta(t)+\langle\rho\rangle\sqrt{\frac{\,k_BT}{\gamma\pi\,t}}\right]$.
The noise $\xi_n(t)$ is assumed to be Gaussian, zero-mean valued and
related to the damping kernel  by Kubo's generalized
fluctuation-dissipation relation \cite{Kubo_1966}, i.e.

\be \langle\xi_n(t)\xi_n(t')\rangle=k_BT\,\kappa(\left|t-t'\right|)
\label{GFDT}. \ee

\noi Because of this property  $\xi_n(t)$ is called fractional
Gaussian noise. The Langevin description of Eqs.
(\ref{GLE})-(\ref{GFDT}) closely reproduces all three diffusive
regimes of the tagged particle and their time-scales: ballistic,
diffusive and subdiffusive (see Fig.\ref{fig1}). The success of this
approach is largely due to the choice of the kernel $\kappa(t)$,
where and {\it ad hoc} $\delta$ term was added to a subdiffusive
long-lasting memory tail. As a matter of fact no rigorous derivation
of the Langevin equations (\ref{GLE})-(\ref{GLE_velocity}) has been
put forward, so far. However, in the overdamped limit, the fractional
terms of the Langevin equation (\ref{GLE}) have been obtained thanks
to a procedure called \emph{harmonization} \cite{Lizana_PRE_2010}.
The resulting fractional Langevin equation for the tagged particle
reads \cite{Mainardi_1996}, \cite{Mainardi_2008}\cite{Lutz_PRE_2001}

\be
2\sqrt{\gamma\,k_BT}\langle\rho\rangle\frac{d^{1/2}x_n(t)}{dt^{1/2}}
=\xi_n(t), \label{FLE} \ee

\noi where
$\langle\xi_n(t)\xi_n(t')\rangle=2\langle\rho\rangle(k_BT)^{3/2}\sqrt{\frac{\gamma}{\pi\,\left|t-t'\right|}}$. We hereby provide just a sketch of the derivation, addressing the interested reader to Ref.s\cite{Lizana_PRE_2010}\cite{Taloni_PRL_2010} for the specific details. The harmonization procedure aims at substituting the hard-core interaction (\ref{potential}) in Eq.(\ref{SF_diffusion}) by its second order expansion around the maximum of the corresponding free energy, hence Eq.(\ref{LE}) transforms, in the overdamped limit, to

\be
\gamma\frac{dx_n(t)}{dt}=\langle\rho\rangle^2k_BT \,[x_{n+1}(t)+x_{n-1}(t)-2x_n(t)]+\eta_n(t).
\label{LE_harmon}
\ee

\noi Passing to the continuum limit $x_n(t)\to x(n,t)$ and introducing the Fourier transform in space and time as $\tilde{x}(q,\omega)=\int_{-\infty}^{\infty}dn\,dt \,x(n,t)e^{-i(qn-\omega t)}$, the previous equation can be solved as

\be
\tilde{x}(q,\omega)=\frac{\tilde{\eta}(q,\omega)}{-i\omega\gamma+\langle\rho\rangle^2k_BTq^2}.
\label{LE_harmon_FT}
\ee

\noi Multiplying both sides by $2(-i\omega)^{1/2}\sqrt{k_BT\gamma}\langle\rho\rangle$ and performing an inverse Fourier transform in the $n$ domain we obtain

\be
2(-i\omega)^{1/2}\sqrt{k_BT\gamma}\langle\rho\rangle\tilde{x}(n,\omega)=\tilde{\xi}(n,\omega)
\label{LE_harmon_FT_2}
\ee

\noi where the fractional Gaussian noise is

\be
\tilde{\xi}(n,\omega)=\gamma\int_{-\infty}^{\infty}dn'\,\tilde{\eta}(n',\omega)e^{-\sqrt{\frac{-i\omega\gamma}{\langle\rho\rangle^2k_BT}}\left|n-n'\right|}.
\label{fGn_FT}
\ee

\noi Inverting back in time, one recovers the Eq.(\ref{FLE}). We stress that the very same equation can be immediately drawn from Eq.\ref{GLE_velocity} by time integration (using Laplace transform) and averaging over the initial velocities.

Besides its compact and elegant notation, the above formalism offers
an additional remarkable advantage: all observables of practical
interest can be calculated analytically. For instance, it becomes
apparent that the subdiffusive law in Eq.(\ref{SF_diffusion}) is
closely related to persistent memory effects, which appear in the
negative power-law tails of the velocity autocorrelation functions
\cite{Percus_PRA_1974}\cite{VanBeijeren_PRB_1983}\cite{Marchesoni_PRL_2006},
\cite{Taloni_PRE_2006}, \cite{Felderhof_JCP_2009}\cite{Tripathi_PRE_2010}: $$\langle
v_n(t)v_n(t')\rangle=-\sqrt{\frac{k_BT}{\gamma\pi}}
\frac{\left|t-t'\right|^{-3/2}}{4\langle\rho\rangle}.$$ Physically,
these long-time anticorrelations originate from the particle momentum
exchange due to the hard-core collisions. By time integrating a
velocity autocorrelation function, one obtains the corresponding
position autocorrelation function, which in the long-time limit can
be cast as \cite{Taloni_PRE_2010}\cite{Taloni_epl_2012}

\be
\langle\left[x_{n}(t)-x_{n}(0)\right]\left[x_{n}(t')-x_{n}(0)\right]\rangle
=\sqrt{\frac{k_BT}{\gamma\pi\langle\rho\rangle^2}}\left[\sqrt{t}+\sqrt{t'}-\sqrt{\left|t-t'\right|}\right].
\label{PAF}
\ee

Now a stochastic process that exhibits persistent correlations, such
those in Eq.(\ref{PAF}), and noise autocorrelations, like
$\langle\xi_n(t)\xi_n(t')\rangle\propto
1/{\sqrt{\left|t-t'\right|}}$, is said to perform a fractional
Brownian motion (FBM) \cite{Mandelbrot_1968}. Hence one could
reasonably conclude that the tagged particle performs a fractional
Brownian motion, which is true, except that in Mandelbrot's original
definition of FBM no mention is made of any (generalized)
fluctuation-dissipation relation. This is a delicate issue which goes
beyond the purpose of the present article. The relation between FLE
and FBM is addressed in Ref.\cite{Lutz_PRE_2001}, while certain
subtleties concerning the validity of the fluctuation-dissipation
relation in the FBM are discussed in Ref.\cite{Taloni_PRL_2010}. 

\noi Yet, another worth mentioning aspect brought forth within the context of the FLE framework  is the emergence of a universality class ~\cite{Taloni_PRE_2008}\cite{Taloni_PRL_2010}. As a matter of fact, the single file model belongs to a class of stochastic systems which includes several linearly interacting models in different area of physics, ranging from  the one dimensional Edward-Wilkinson chain \cite{Edwards_Proc_Roy_Soc_Lon_1982}\cite{Bustingorry_PRE_2010}, to fluctuating interfaces \cite{Toroczkai_Phys_today_2008}, to Rouse polymers \cite{Rouse_JCP_2004}. In these many-body systems indeed, a tagged probe undergoes an anoumalous diffusion whose asymptotic behaviour is governed by the Eq.(\ref{FLE}). Most importantly, the dynamics of the distance between the donor (\emph{D}) and the acceptor (\emph{A}) coordinates within a protein strain was shown to be reproduced, with an excellent degree  of accuracy, by a FLE with an applied hookean force  ~\cite{Kou_PRL_2004}. In Ref.\cite{Lizana_PRE_2010} it has been rigourously proven  that the distance between two particles in single file systems, $\Delta_{n,n'}(t)=x_n(t)-x_{n'}(t)$, asymptotically recovers the FLE for the donor-acceptor distance (see also \cite{Taloni_PRE_2010}). Strictly speaking, the quantity  $\Delta_{n,n'}(t)=x_n(t)-x_{n'}(t)$ behaves as a stochastic subdiffusive trajectory in an harmonic well, whose frequency is proportional to $\sqrt{n-n'}$. Rephrasing this statement, two particles $n$ and $n'$ are \emph{de facto} connected by an entropic spring with stiffness equal to $\left|n-n'\right|$.

We stress once more that, on jumping  from the full file dynamics of
Eq.(\ref{LE}) to the reduced GLE-FLE's (\ref{GLE})-(\ref{FLE}) of the
tagged particle, no systematic projection procedure was employed to
eliminate the coordinates of the untagged file particles. Indeed, a
controllable ``adiabatic elimination'' procedure of the fast
variables, for instance, along the line of the Mori-Zwanzig formalism
~\cite{Mori_1956}\cite{Mori_1965_first}\cite{Mori_1965_second}\cite{Zwanzig_1961},
cannot be carried out for a single file in the real space (as opposite to the Fourier space), where the time scales of all
particle coordinates are identical. How are then the spatial file
interactions of Eq.(\ref{LE}) accounted for in the phenomenological
GLE formalism, Eqs. (\ref{GLE})-(\ref{FLE})? They partially translate
into the time correlations contained in the definition of the
fractional derivative, see Eq.(\ref{caputo_der}), and in the
spatio-temporal properties of the fractional Gaussian noise. Such a
mechanism is apparent in the overdamped limit, where the
autocorrelation of the fractional noise reads \cite{Taloni_PRE_2008}\cite{Taloni_PRL_2010}

\begin{equation}
\begin{array}{l}
\langle\xi_n(t)\xi_{n'}(t')\rangle
=\frac{(2k_BT)^{3/2}\sqrt{\gamma}}{\pi}\,\int_0^{\infty}d\omega
\frac{e^{-\frac{\left|n-n'\right|}{\langle\rho\rangle}\sqrt{\frac{\omega}{2\gamma}}}}{\sqrt{\omega}}\\
\cos\left(\omega\left|t-t'\right|\right)\left[\cos\left(\frac{\left|n-n'\right|}{\langle\rho\rangle}
\sqrt{\frac{\omega}{2\gamma}}\right)+\sin\left(\frac{\left|n-n'\right|}{\langle\rho\rangle}\sqrt{\frac{\omega}{2\gamma}}\right)\right].
\label{GFDT_2}
\end{array}
\end{equation}

\noi Note that for $n=n'$ this expression of the noise
autocorrelation coincides with Eq.(\ref{GFDT}). By making use of this
property when taking the time integration of the FLE (\ref{FLE}), one
can easily calculate the long-time behavior of some particle-particle
correlation functions, such as
$\langle\left[x_{n}(t)-x_{n}(0)\right]\left[x_{n'}(t')-x_{n'}(0)\right]\rangle$
\cite{Kumar_PRE_2008}\cite{Taloni_PRE_2010} or
$\langle\left[x_{n}(t)-x_{n}(0)\right]v_{n'}(t')\rangle$
\cite{Taloni_PRE_2008}.

We now discuss how Eqs.(\ref{GLE})-(\ref{FLE}) are modified by the
presence of the interacting potential $U_{\rm int}$.

\begin{itemize}

\item \emph{File interaction with a substrate.}  In this case no GLE
for a tagged particle was ever proposed, not even in the overdamped
limit. In particular, the question whether the harmonization
technique can be extended to handle the case of a single file
interacting with an underlying substrate, is still unanswered.

\item \emph{One-particle interaction}. When the external perturbation acts
upon the tagged particle alone, an analytical derivation of the GLE
for all file particles is still doable, at least in the overdamped
limit. However, the resulting FLE will be different depending on
whether one considers the tagged or untagged particles. In
Ref.\cite{Lizana_PRE_2010} the FLE of the tagged particle was derived
for the case of a constant force. A general formulation valid for any
kind of one-particle potentials was obtained in
Ref.\cite{Taloni_PRE_2011}. For the tagged particle the reduced FLE
turned out to be

\be
2\sqrt{\gamma\,k_BT}\langle\rho\rangle\frac{d^{1/2}x_n(t)}{dt^{1/2}}
= -\frac{\partial}{\partial
  x_n} U(x_n,t) + \xi_n(t),
\label{FLE_tagged_part_applied_potential}
\ee

\noi where $U(x_n,t)=\int dX\,U(X;t)\delta(X-x_n)$, like in
Eq.(\ref{int_singlepart}). More remarkably, the FLE of an untagged
particle of label $n'$,

\begin{equation}
\begin{array}{l}
2\sqrt{\gamma\,k_BT}\langle\rho\rangle\frac{d^{1/2}x_{n'(t)}}{dt^{1/2}}=\\
 -\frac{\partial}{\partial
  x_n}\int_{-\infty}^{t} dt'\,U(x_n,t')\Theta\left(\left|n-n'\right|, t-t'\right) +
  \xi_{n'}(t),
\label{FLE_untagged_part_applied_potential}
\end{array}
\end{equation}

\noi was shown to depend on the force propagator $\Theta$, i.e. the space-time Green's function which is carrying the external perturbation, while exerted on particle $n$ at time $t'$, to the particle $n'$ at time $t$ \cite{Taloni_PRE_2011}. Indeed,
although applied to the sole tagged particle, $n$, the perturbation
propagates through the file via the hard-core collisions, until it
finally reaches the untagged particle, $n'$. The expression of
$\Theta$ was obtained in terms of its time Fourier transform, that is

\be
\Theta\left(\left|n-n'\right|, \omega\right)=\sqrt{\frac{\langle\rho\rangle}{\gamma}}
(k_BT)^{1/4}e^{-\frac{\left|n-n'\right|}{\langle\rho\rangle}\sqrt{\frac{-i\omega}{\gamma}}}.
\label{theta}
\ee

\noi We remark that the noise appearing in
Eqs.(\ref{FLE_tagged_part_applied_potential}) and
(\ref{FLE_untagged_part_applied_potential}) satisfies the same
properties as the noise in Eq.(\ref{FLE}).

\item \emph{File particle-particle interaction}. In the presence of binary interactions
between nearest-neighbors, a tagged-particle FLE has been rigorously
derived only for the quadratic interaction of Eq.(\ref{potential_LR})
\cite{Taloni-unpublished}. In the long-time limit, the relevant FLE
for $1<z<2$ boils down to

\be
z\langle\rho\rangle\sin\left(\frac{\pi}{z}\right)k_z^{1/z}\gamma^{\frac{z-1}{z}}\frac{d^{\frac{z-1}{z}}x_n(t)}{dt^{\frac{z-1}{z}}}
= \xi_n(t), \label{FLE_part_int} \ee

\noi where the fractional Gaussian noise satisfies the generalized
fluctuation-dissipation relation
$$\langle\xi_n(t)\xi_n(t')\rangle=\frac{k_BTz\langle\rho\rangle\sin\left(\frac{\pi}{z}\right)k_z^{1/z}
\gamma^{\frac{z-1}{z}}}{\Gamma\left(1/z\right)\left|t-t'\right|^{\frac{z-1}{z}}}.$$

\end{itemize}

\section{From the diffusion-noise equation to the fractional Langevin equation}

We address now the single file dynamics from a different viewpoint.
Let us start from the Langevin equations (\ref{LE}) of a
noninteracting single file with $U_{\rm int}=0$; in
Eq.(\ref{potential}) only the hard-core collision term, $U_{HC}$, is
retained. We then concentrate on the time evolution of the particle
density rather than on the particle dynamics. To do so, let us divide
the ring in $M$ bins of length $\Delta X$ and define the coarse
grained particle density

\be \rho_i(t)=\frac{1}{\Delta
X}\sum_{n=1}^{N}\vartheta\left[x_n(t)-(i-1)\Delta X\right]
\vartheta\left[i\Delta X-x_n(t)\right], \label{dscrt_density} \ee

\noi where $i$ is the bin label, $\vartheta$ denotes the Heaviside
step function, and $\Delta X=L/M$. As we are interested in the
thermodynamic limit, where the file equilibrium is characterized by
the constant density $\langle\rho\rangle=\lim_{N\to \infty, L\to\infty} N/L$,
the bin size can be taken arbitrarily small, $\Delta X\to 0$.
Note that by virtue of the $\vartheta$ definition,
$\vartheta[X]=\int_{-\infty}^Xds\, \delta(s)$, and  the approximation
$\int_{(i-1)\Delta X}^{i\Delta
X}ds\,\delta\left(s-x_n(t)\right)\simeq \delta\left((i-1/2)\Delta
X-x_n(t)\right)\Delta X$, the standard file density,

\be \rho_i(t)\to\rho(X,t)=\sum_{n=1}^N \delta\left(X-x_n(t)\right),
\label{ct_density} \ee

\noi can be immediately recovered.

Describing the single file in terms of the coarse grained density
surely implies a loss of information. Nevertheless, one can study the
stochastic behavior of the bin density, $\rho_i(t)$, and try to
relate its fluctuations to the tagged particle dynamics. To this
purpose, we first notice that, being a local property of the file,
the density definition in Eq.(\ref{dscrt_density}) does not depend on
the particle relabeling upon collisions; differently stated, the
coarse grained density of a single file does not change if we replace
the non-passing file particles with noninteracting Brownian particles
\cite{Lebowitz_PR_1967}. We recall that in the absence of hard-core
collisions the time evolution of the density profile is governed by
the continuity equation \cite{VanVliet_JMath_Phys_1971}

\be \frac{\partial}{\partial t} \delta\rho(X,t) =
-\frac{\partial}{\partial X} J(X,t), \label{DN} \ee

\noi where the density function has been separated into a noise
average, $\langle\rho(X,t)\rangle$, and a fluctuating part,
$\delta\rho(X,t)$, that is,
$\rho(X,t)=\langle\rho(X,t)\rangle+\delta\rho(X,t)$. The density
current appearing on the r.h.s. of Eq.(\ref{DN}) also evolves
according to a stochastic equation,

\be J(X,t) =-D\frac{\partial}{\partial X}\delta\rho(X,t)+\zeta(X,t),
\label{flux} \ee

\noi with a noise term, $\zeta(X,t)$, which satisfies the following
identities

\be
\begin{array}{l}
\zeta(L,t)=\zeta(0,t),\\
\langle \zeta(X,t)\rangle=0,\\
\langle
\zeta(X,t)\zeta(X',t')\rangle=2D\delta(X-X')\delta(t-t')\langle\rho(X,t)\rangle.
\label{noiseproperty}
\end{array}
\ee

\noi The periodic boundary condition formulated in the first identity
is required to ensure the conservation of the particle number in the
segment (or ring) $[0,L]$ (\emph{conserved noise})
\cite{Pototsky_PRE_2010}. The function $\langle\rho(X,t)\rangle$
encodes the spatio-temporal evolution of the noise-averaged density
for given initial conditions, $\rho(X,0)$. In particular, a uniform
particle distribution at time $t$ corresponds to setting
$\langle\rho(X,t)\rangle\equiv\langle\rho\rangle$. The set of
Eqs.(\ref{DN}) and (\ref{flux}) is generally referred to as the
\emph{diffusion-noise equation} \cite{Van_Kampen}. We also recall that Eq.(\ref{DN}) can be seen as the linear approximation of the Dean-Kawasaki equation \cite{Dean_JPhysA_1996}, which constitutes the exact evolution equation for the microscopic density of a collection of interacting Brownian particles under overdamped dynamics \cite{Kim_PRE_2014}.

The first authors who related the coarse grained density fluctuations
to the mean square displacement of the tagged particles, were
Alexander and Pincus \cite{Alexander_PRB_1978}. Thirty years later,
another approach, still based on the density evolution equation
(\ref{DN}), was proposed to obtain the same results, but under
different approximations \cite{Taloni_PRE_2008}. We hereby review
both approaches and show that they are indeed complementary. Before proceeding further, however, we mention two papers recently appeared on the connection between tagged particle dynamics and a liquid-theory approach at the mesoscopic scale, i.e. a Lagrangian formulation for the file density evolution \cite{goto_JPSJ_2011}\cite{Takeshi_PRE_2013}. However it must be stressed that the above formalism holds for a system of interacting particles, while the present approach exploits the connection of an assembly of non-interacting Brownian particles and single file system. 

On introducing the space Fourier transform of $\rho(X,t)$,
$\tilde{\rho}(Q,t)=\int_{-\infty}^{\infty}dX\,e^{-iQX}\rho(X,t)$, we
rewrite Eq.(\ref{ct_density}) as

\be
\rho(X,t)=\int_{-\infty}^{\infty}dQ\,\sum_{n=1}^Ne^{-iQ\left[x_n(t)-X\right]},
\label{DN_1}
\ee

\noi which, in turn, yields

\be
\tilde{\rho}(Q,t)=\sum_{n=1}^Ne^{-iQx_n(t)}.
\label{DN_2}
\ee

\noi Now, we write the trajectory of a tagged (non-passing) file
particle as $x_n(t)=\frac{n}{\langle \rho\rangle}+\delta x_n(t)$,
upon assuming that the motion of the $n$-th particle takes place
around its equilibrium position, $\langle x_n(t)\rangle
=\frac{n}{\langle \rho\rangle}$. Accordingly,

\be \tilde{\rho}(Q,t)\simeq\sum_{n=1}^Ne^{-iQ\frac{n}{\langle
\rho\rangle}}-iQ\sum_{n=1}^N\delta x_n(t)e^{-iQ\frac{n}{\langle
\rho\rangle}}, \label{DN_3} \ee

\noi where deviations from the average particle position have been
treated as perturbatively small, $\delta x_n(t)\simeq 0$ ($\delta x_n(t)\ll Q^{-1}$).
The first term on the r.h.s. of Eq.(\ref{DN_3}) is the inverse
Fourier transform of $\langle \rho\rangle$; hence,

\be
\delta\tilde{\rho}(Q,t)\simeq-iQ\sum_{n=1}^N\delta x_n(t)e^{-iQ\frac{n}{\langle \rho\rangle}}.
\label{DN_4}
\ee

\noi On introducing the discrete Fourier transform of the particle
trajectory, $\tilde{x}_k(t)=\sum_{n=1}^N x_n(t)e^{-i\frac{2\pi
k}{N}n}$ (with $k$ integer wavenumbers $k=0,\pm1, \pm 2, \dots$), simple algebraic passages lead to the Alexander and Pincus
relation connecting density and particle fluctuations, namely,

\be
\delta\tilde{\rho}(Q,t)\simeq-iQ\left.\delta\tilde{x}_k(t)\right|_{k=\frac{NQ}{2\pi\langle \rho\rangle}}.
\label{DN_5}
\ee

\noi To obtain the mean square displacement of the tagged particle
under stationary conditions, it suffices now to note that $\langle
\left[x_{n}(t)-x_{n}(0)\right]^2\rangle=2\left[\langle\delta
x_n^2(0)\rangle-\langle\delta x_n(t)\delta x_n(0)\rangle\right]$.
Moreover, the autocorrelation function $\langle\delta x_n(t_1)\delta
x_n(t_2)\rangle$ can be readily calculated by means of
Eq.(\ref{DN_5}). Such a task is greatly simplified in the continuum
space limit, $x_n(t)\to x(n,t)$ and
$\tilde{x}(q,t)=\int_{-\infty}^{\infty}dn \,x(n,t)e^{-iqn}$, with $q=\frac{2\pi k}{N}$. On
inverting Eq.(\ref{DN_5}), one eventually obtains the identity

\begin{eqnarray} &&\langle\delta \tilde{x}(q_1,t_1)\delta \tilde{x}(q_2,t_2)
\rangle\\ \nonumber &\simeq& \left.-\frac{\langle\delta
\tilde{\rho}(Q_1,t_1)\delta
\tilde{\rho}(Q_2,t_2)\rangle}{Q_1Q_2}\right|_{Q_1=q_1\langle\rho\rangle,Q_2=q_2\langle\rho\rangle},
\label{DN_6} \end{eqnarray}

\noi whose r.h.s. can be evaluated in the framework of the
diffusion-noise equation,  Eqs.(\ref{DN}) and (\ref{flux}), by making
explicit use of the noise properties in Eq.(\ref{noiseproperty}). The
final result is the autocorrelation function of the particle
fluctuations \cite{Alexander_PRB_1978}

\be
\langle\delta x(n,t_1)\delta x(n,t_2)\rangle\simeq \int_0^{\infty}
\frac{dq_1}{\pi}\frac{e^{-D\langle\rho\rangle^2q_1^2\left|t_1-t_2\right|}}{\langle\rho\rangle^2q_1^2}.
\label{DN_7}
\ee

\noi The asymptotic anomalous diffusion law of
Eq.(\ref{SF_diffusion}) can be easily recovered by rewriting
Eq.(\ref{DN_7}) as

\be
\langle\left[x(n,t)-x(n,0)\right]^2\rangle\simeq\frac{2}{\langle\rho\rangle^2}
\int_0^{\infty}\frac{dq_1}{\pi}\frac{1-e^{-D\langle\rho\rangle^2q_1^2t}}{q_1^2},
\label{DN_8} \ee

\noi and explicitly calculating the integral there.

The alternative approach of Ref.\cite{Taloni_PRE_2008} hinges on an
explicit relation connecting particle fluctuations and current
density. Such a relation can be derived immediately within the Alexander's and Pincus' former derivation. Indeed, by combining Eq.(\ref{DN_5}) and the Fourier transform of
Eq.(\ref{DN}) in the $q$ domain, one has

\be \delta \tilde{x}(q,t)\simeq\int_0^t dt'
\left.\tilde{J}(Q,t')\right|_{Q=q\langle\rho\rangle}. \label{DN_9} \ee

\noi By taking the inverse Fourier transform in the coarse grained
$x$ domain ($n$ domain), it is
obtained the  approximate relation \cite{Taloni_PRE_2008}

\be
\frac{d}{dt}\,\delta x(n,t)\simeq\frac{ J\left(\frac{n}{\langle\rho\rangle},t\right)}{\langle\rho\rangle},
\label{DN_10}
\ee

\noi from which the integral expression (\ref{DN_8}) and the SFD law
(\ref{SF_diffusion}) follow suit. Although slightly different, Taloni and Lomholt approach provides additional
insight into the single file dynamics. Indeed, their starting point
was the coarse graining integral identity,

\be \int_{x(n,t)}^{x(n',t)}dX\,\rho(X,t)=n'-n, \label{DN_11} \ee

\noi which is equivalent, upon time derivation of both sides, to the
spatio-temporal differential relation,

\begin{eqnarray}
&&
\frac{dx(n,t)}{dt}\rho(x(n,t),t)-\frac{dx(n',t)}{dt}\rho(x(n',t),t)
\\
\nonumber &=&-\int_{x(n,t)}^{x(n',t)}dX\,\frac{d}{dt}\rho(X,t),
\label{DN_12}
\end{eqnarray}

\noi and, in view of the continuity equation (\ref{DN}), to the much
simpler equality,

\be \frac{dx(n,t)}{dt}\,\rho(x(n,t),t)=J\left(x(n,t),t\right),
\label{DN_13} \ee

\noi which holds for any $n$ and $n'$. In a consistent coarse
graining formalism, Eq.(\ref{DN_10}) is thus recovered by imposing
$\rho(x(n,t),t)\simeq \langle\rho\rangle$ and
$J\left(x(n,t),t\right)\simeq
J\left(\frac{n}{\langle\rho\rangle},t\right)$.

We take a step forward by writing the time Fourier
transform of Eq.(\ref{DN_9}) as

\be \delta\tilde{x}(q,\omega)\simeq
\left.\frac{\tilde{\zeta}(Q,\omega)}{DQ^2-i\omega}\right|_{Q=q\langle\rho\rangle},
\label{DN_14} \ee

\noi  where we again made use of Eqs.(\ref{DN}) and (\ref{flux})
and defined $\tilde{x}(q,\omega)=\int_{-\infty}^{\infty}dt
\,\tilde{x}(q,t)e^{-i\omega t}$. The corresponding inverse Fourier
transform in the $n$-domain is

\be
\delta\tilde{x}(n,\omega)\simeq\int_{-\infty}^{\infty}dX\,\tilde{\zeta}(X,\omega)
\frac{e^{-\left|\frac{n}{\langle\rho\rangle}-X\right|\sqrt{\frac{-i\omega}{D}}}}{2\sqrt{-i\omega
D}}, \label{DN_15} \ee

\noi or, equivalently,

\be
2\sqrt{k_BT\gamma}\sqrt{-i\omega}\,\delta\tilde{x}(n,\omega)\simeq
\int_{-\infty}^{\infty}dX\,\tilde{\zeta}(X,\omega)\gamma e^{-\left|\frac{n}{\langle\rho\rangle}-X\right|\sqrt{\frac{-i\omega}{D}}}
\label{DN_16}.
\ee

\noi On further Fourier transforming back to the time domain, they
finally recovered the FLE (\ref{FLE}) with fractional time derivative
$\frac{d^{1/2}}{dt^{1/2}}\delta
x(n,t)=\int_{-\infty}^{\infty}\frac{d\omega}{2\pi}
\sqrt{-i\omega}\,\delta\tilde{x}(n,\omega)e^{i\omega t}$
\cite{Podlubny} and fractional Gaussian noise
$\xi(n,t)=\int_{-\infty}^{\infty}\frac{d\omega}{2\pi}\int_{-\infty}^{\infty}dX\,\tilde{\zeta}(X,\omega)\gamma
e^{-\left|\frac{n}{\langle\rho\rangle}-X\right|\sqrt{\frac{-i\omega}{D}}+i\omega t}$ (see Eq.(\ref{fGn_FT}))
\cite{Lizana_PRE_2010,Taloni_PRL_2010}.

Finally, we notice that the harmonization technique could be extended to the interaction with a substrate within he diffusion-noise approach, following the Dean's oute traced for a system of interacting Brownian particles.

\section{Conclusions}

In this paper we reviewed select advances in the field of single file
diffusion. Our attention focused mostly on the effective stochastic
equation that governs the motion of a single file particle, the
so-called tagged particle, when the time evolution of the remaining
file particles is ignored or not accessible. Such an equation allows
the direct analytical calculation of all correlation functions and,
therefore, of all relevant physical observables. For instance,
dynamical structure factors and file transport properties ultimately
depend on the asymptotic diffusion law of the tagged particle.
Moreover, the generalized Langevin framework encompasses the presence
of external perturbations in a very intuitive and analytically
tractable form. Changing perspective, we have shown how the
fluctuations of the density profile along the substrate are strictly
related to the random motion of tagged particle. This relation,
obtained through two different complementary approaches, sheds light
on the intimate connection existing between tagged particle and
density fluctuations and, equivalently, between fractional Langevin
equation and diffusion-noise equation formalism. This constitutes a
truly powerful theoretical tool,  allowing  an observer to infer the
diffusion properties of a single particle from the spatio-temporal
sampling of the overall file density.

\bibliographystyle{ws-brl}
\bibliography{SFS}

\end{document}